\documentclass[%
 reprint,
 amsmath,amssymb,
 longbibliography,
 aps,
]{revtex4-1}
   \usepackage[pdftex]{graphicx}
\usepackage{amsmath}
\usepackage{tabularx}
\usepackage{url}
\usepackage{soul,color}

\begin{document}
%
\preprint{APS/123-QED}
\title{A Photonic In-Memory Computing primitive for Spiking Neural Networks using Phase-Change Materials}
%
%
%

\author{Indranil Chakraborty}
\email{ichakra@purdue.edu}
\author{Gobinda Saha}%
 \author{Kaushik Roy}%

\affiliation{%
 School of Electrical \& Computer Engineering, Purdue University, West Lafayette, IN 47907, USA
\\
}

\date{\today}

\begin{abstract}
Spiking Neural Networks (SNNs) offer an event-driven and more biologically realistic alternative to standard Artificial Neural Networks based on analog information processing. This can potentially enable energy-efficient hardware implementations of neuromorphic systems which emulate the functional units of the brain, namely, neurons and synapses. Recent demonstrations of ultra-fast photonic computing devices based on phase-change materials (PCMs) show promise of addressing limitations of electrically driven neuromorphic systems. However, scaling these standalone computing devices to a parallel in-memory computing primitive is a challenge. In this work, we utilize the optical properties of the PCM, Ge\textsubscript{2}Sb\textsubscript{2}Te\textsubscript{5} (GST), to propose a Photonic Spiking Neural Network computing primitive, comprising of a non-volatile synaptic array integrated seamlessly with previously explored `integrate-and-fire' neurons. The proposed design realizes an `in-memory' computing platform that leverages the inherent parallelism of wavelength-division-multiplexing (WDM). We show that the proposed computing platform can be used to emulate a SNN inferencing engine for image classification tasks. The proposed design not only bridges the gap between isolated computing devices and parallel large-scale implementation, but also paves the way for ultra-fast computing and localized on-chip learning.  
\end{abstract} 
\maketitle

%

\section{Introduction}
%
%
%
%
The phenomenal success in the field of Deep Learning using Artifical Neural Networks (ANN) based on analog information processing has had far reaching consequences in the past decade \cite{LeCun_2015}. Machines driven by such networks have surpassed human in various tasks ranging from pattern recognitions to playing complex games such as Go \cite{Silver_2016} and Chess \cite{Campbell_2002}. However, the growing complexities of computational models involved in such multi-layered neural networks have rendered the training and inferencing tasks extremely expensive in terms of memory and energy. The gulf between the energy efficiency of the brain and standard neural network architectures have led researchers to explore a bio-plausible alternative, namely, Spiking Neural Networks (SNNs). The event-driven nature and sparse information encoding of SNNs make them more feasible for energy-efficient neuromorphic computing thus paving the way towards unraveling the elusiveness of the brain. The fundamental operations performed by SNNs involve parallelized dot-product through the synaptic network followed by subsequent integration and thresholding by the neurons. Neuromorphic systems attempting to leverage the sparse and event-driven nature of SNNs thus aim toward efficient emulation of these functionalities. \par
The initial efforts \cite{Serrano_Gotarredona_2009,Merolla_2014,Furber_2014} in hardware implementations of SNNs was based on standard von-Neumann architecture \cite{von2012computer} based on Complementary Metal Oxide Seminconductor (CMOS) technology where the synaptic units of the neural networks are stored in the digital memory and repeatedly fetched by the processor for computing operations. However, the overhead of frequent data transport between the memory and processor have led to a shift in the computing paradigm as `in-memory' computing platforms \cite{Biswas_2018,jaiswal20188t} attempt to emulate the `massively parallel' operations of the brain. Although the term `neuromorphic' was primarily coined \cite{Mead_1990} with CMOS technology in mind, this computing domain has branched out to non-volatile memory (NVM) technologies such as oxide-based memristors \cite{Li_2017}, spintronics \cite{sengupta2017encoding}, phase change materials (PCM) \cite{Eryilmaz_2014,Tuma_2016}, etc in the recent years. The natural ability of these resistive technologies to compute parallelized dot-products using crossbar structures make them promising candidates for neuromorphic systems. Despite the extensive efforts in NVM-based in-memory computing in the electrical domain, these technologies suffer from different drawbacks manifesting in form of energy-efficiency, speed and sneak paths. Moreover,  write latencies into memristors \cite{rajendran2013specifications,shafiee2016isaac}  is  a  major  reason  why  memristive  devices  are  not  suitable  for  temporally  scalable  architectures. Thus, there  is  a  need  to  explore  a  different  memory  technology  which  can  enable  computing  as  well  as  the  possibility of  lower  write  times. \par
Integrated Photonics offers an alternative approach to standard microelectronic `in-memory' computing platforms and promises ultra-fast neural computing and information processing. The recent advances in photonics-based neuromorphic computing has overseen implementations of various kinds \cite{Vandoorne_2014,shen2017deep} of neural processing units on the photonic platform leveraging the inherent capability of matrix operations of integrated optical circuits. Spike-based processing systems have also been extensively explored using excitable lasers \cite{tait2014broadcast,Tait_2017}. However, most of the photonic systems investigated in the context of neuromorphic computing are based on volatile information processing which require thermal tuners to maintain the modulation states which might turn out to be energy expensive for large-scale systems. Non-volatility offers the ability to write and erase information dynamically desirable for large-scale implementations of neuromorphic systems. To that effect, recent demonstrations of sub-ns writing speeds in GST-based PCM technology through optical pulses has opened up a host of opportunities of in-memory computing in the photonic domain\cite{rios2018memory}. The ultra-fast switching using light overcomes the longstanding obstacle of high `write' latencies \cite{rajendran2013specifications} for PCMs in the electrical domain. The highly contrasting optical properties of GST in its crystalline and amorphous phases have led to implementations of all-photonic memories \cite{Rios_2015}, switches \cite{Stegmaier_2016} and reconfigurable non-volatile computing platforms \cite{Zheng_2018}. More recently, photonics-based GST devices have also been explored to emulate biologically plausible synapses \cite{cheng2017chip}, capable of undergoing Spike Timing Dependent Plasticity (STDP), and `integrate and fire' spiking neurons \cite{chakraborty2018all}.  Despite these promising investigations towards fast neural computing based on non-volatile platform, the challenge of scaling standalone devices to large-scale neuromorphic systems is enormous. Thus, there is a need to explore non-volatile memory primitive in the photonic domain, which can perform parallel computing. In this work, we propose an all-photonic SNN computing primitive, based on GST-based photonic neural elements, which attempts to bridge the gap between devices to system-level implementation of Photonic neural networks. We leverage the inherent wavelength division multiplexing (WDM) \cite{yang2012chip} property of optical networks to propose a non-volatile synaptic array, while exploring and mitigating the challenges arising from designs based on ring resonators of radii comparable to the wavelength of operation. Such a synaptic array can achieve higher densities compared to current state-of-art photonic computing systems. We show how the proposed synaptic computing platform can be seamlessly integrated with previously explored `integrate and fire' spiking neurons to realize an ultra-fast and truly integrable Spiking Neural Network. Finally, we evaluate the performance of the proposed Photonic SNN in the classification task of handwritten digits. 

\section{Photonic Synapses}
The core computational units of any neural network are neurons and synapses. In SNNs, information is encoded in form of spikes and the neurons and synapses are capable of processing information through these spike trains. As shown in Fig. \ref{fig:Syndev} (a), the input trains of spikes get multiplied by the synaptic weights $w_1, w_2, ..., w_n$ and the weighted sum is received by an `Integrate-and-Fire' neuron. The internal state of the neuron, known as the `membrane potential' ($V_{mem}$) integrates based on the incoming weighted spikes and is compared with a threshold ($V_{th}$) at every time-step. The neuron outputs a spike once $V_{mem}$ reaches $V_{th}$. The synaptic functionality essentially corresponds to a multiplication operation of the inputs and the corresponding weights of the synapses. The basic operation performed by a single synapse can be represented as $I_iw_i$. We show how a single bus microring resonator with a GST element embedded on top of it can operate as such a synapse. The device under consideration is a Si-on-insulator structure consisting of a rectangular waveguide and a ring waveguide as shown in Fig. \ref{fig:Syndev} (b). A GST element is deposited on one arm of the ring waveguide, which takes the shape of an arc and the length of the arc is denoted as the length of the GST element ($L_{GST}$). The fabrication technique of building such a structure has been well explored \cite{Stegmaier_2016,Zheng_2018}. Wave in the rectangular waveguide gets partially coupled to the ring and constructively interferes when the round-trip phase shift equals an integer multiple of $2\pi$ leading to the resonant condition:
\begin{figure}[t]
		\centering
		\includegraphics[width=3.6in, keepaspectratio]{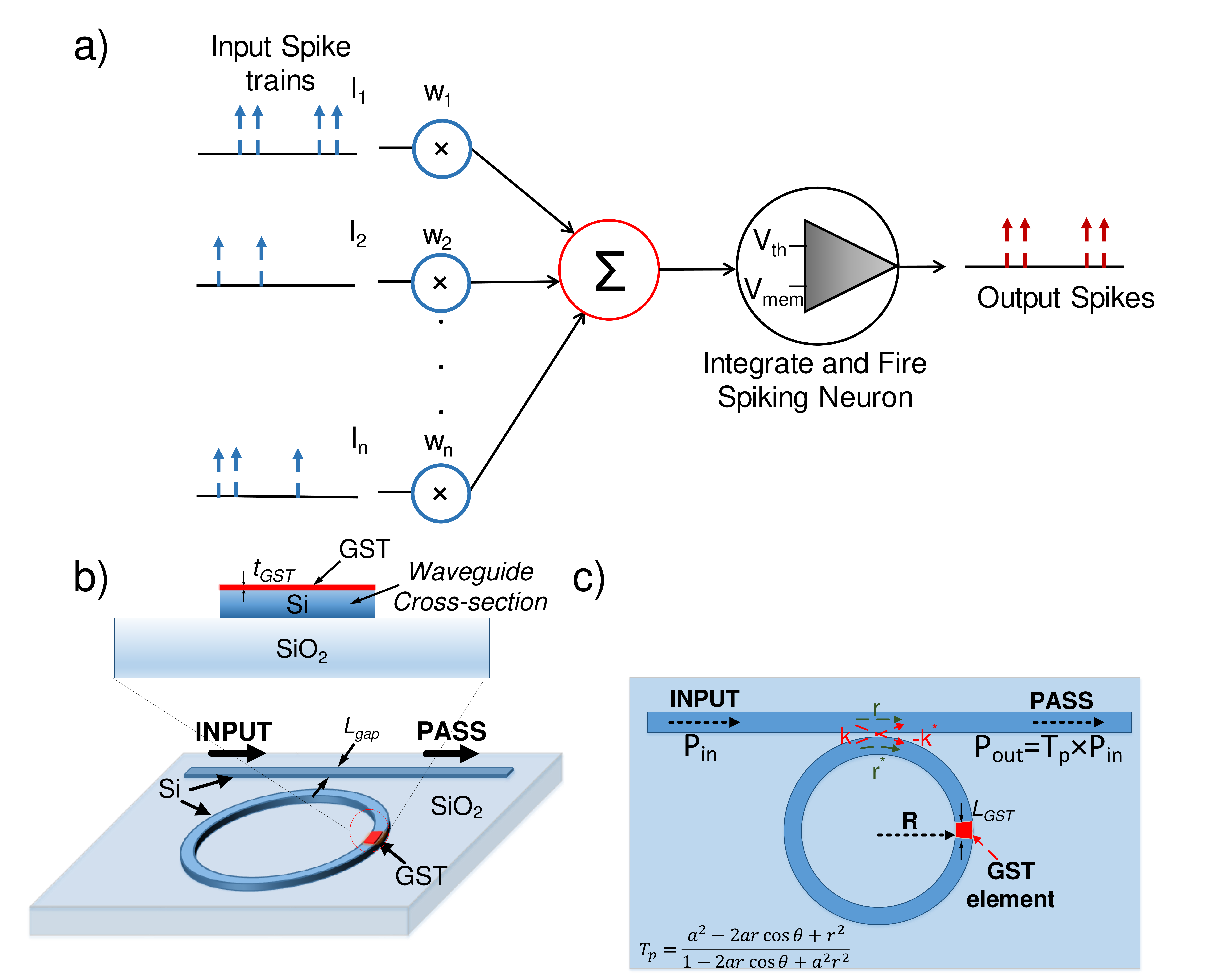}
		\caption{(a) The basic functional elements of an SNN are spiking neurons and weighted synaptic connections. At each time instant, the inputs are weighted by the synaptic weights to produce a resultant output represented as $\sum_iP_iw_i$. The `integrate-and-fire' neuron's membrane potential ($V_{mem}$) is updated according to the weighted sum and compared with a threshold value ($V_{th}$). (b) GST-embedded single bus microring resonator structure with Si waveguides on SiO\textsubscript{2} substrate. (c) Top view of the device illustrating the different parameters pertaining to the ring resonator structure. The synaptic device performs an analog multiplication of input $P_{in}$ and transmission $T$.}
		\label{fig:Syndev}
	\end{figure}
\begin{equation}
2\pi R_{ring}n_{eff,wg} = m\lambda_m
\end{equation}
where $R_{ring}$ is the radius of the ring waveguide, $n_{eff,wg}$ is the effective refractive index of the ring waveguide and $\lambda_m$ is the resonant wavelength. The transmission through the `PASS' port is dependent on the device dimensions and material such that:
\begin{equation}
T_{p} = \frac{a^2-2arcos\theta+r^2}{1-2arcos\theta+a^2r^2}
\end{equation}
where $a$ is the attenuation factor and $r$ is the self-coupling coefficient as shown in Fig. \ref{fig:Syndev} (c). $\theta$ is the single-pass phase shift. Under resonance, $\theta$ equals $2\pi$ and the transmission is given by $T_{min} = ((a-r)/(1-ar))^2$. \par

We leverage the contrasting optical properties of GST in its amorphous (a-GST) and crystalline (c-GST) states to manipulate the attenuation in the ring waveguide and thus vary the transmission $T_{min}$ at the resonance wavelength. The varying imaginary refractive indices of a-GST and c-GST leads to differential absorption of evanescently coupled light. The difference in optical absorption can be visibly observed through the cross-section view of the fundamental mode profiles in GST-embedded Si waveguide when excited by a TE mode electromagnetic (EM) wave as shown in Fig. \ref{fig:mode}. c-GST introduces a significant change in waveguide mode in contrast to a-GST due to higher absorption in the GST element. The attenuation factor ($a$) in Eqn. 2 can be related to the imaginary refractive index as: 
\begin{equation}
a = exp(-\frac{2\pi\kappa_{eff,GST}L_{GST}}{\lambda}+Loss)
\end{equation}
where $\kappa_{eff,GST}$ is the effective imaginary refractive index of the GST on Si-SiO\textsubscript{2} stack, $L_{GST}$ is the length of the GST element, and the term `$Loss$' refers to other propagation losses such as bending losses, etc. The GST element can be programmed to partially crystallized levels such that multi-level states can be achieved \cite{Rios_2015,Zheng_2018}. To note, from the perspective of neural networks, significant progress have been made towards proposing training algorithms \cite{hubara2016binarized, Rastegari_2016} which preserve performance even with binarized synapses. Thus, although multi-level states would be desirable from a device point of view, modified training techniques can enable reasonable performance with low-precision synapses. 
\par
The refractive indices of partially crystallized GST can be calculated from effective permittivities approximated by an effective-medium theory \cite{Chen_2015,Voshchinnikov_2007}:
\begin{figure}[t]
        
		\centering
		\includegraphics[width=3.6in, keepaspectratio]{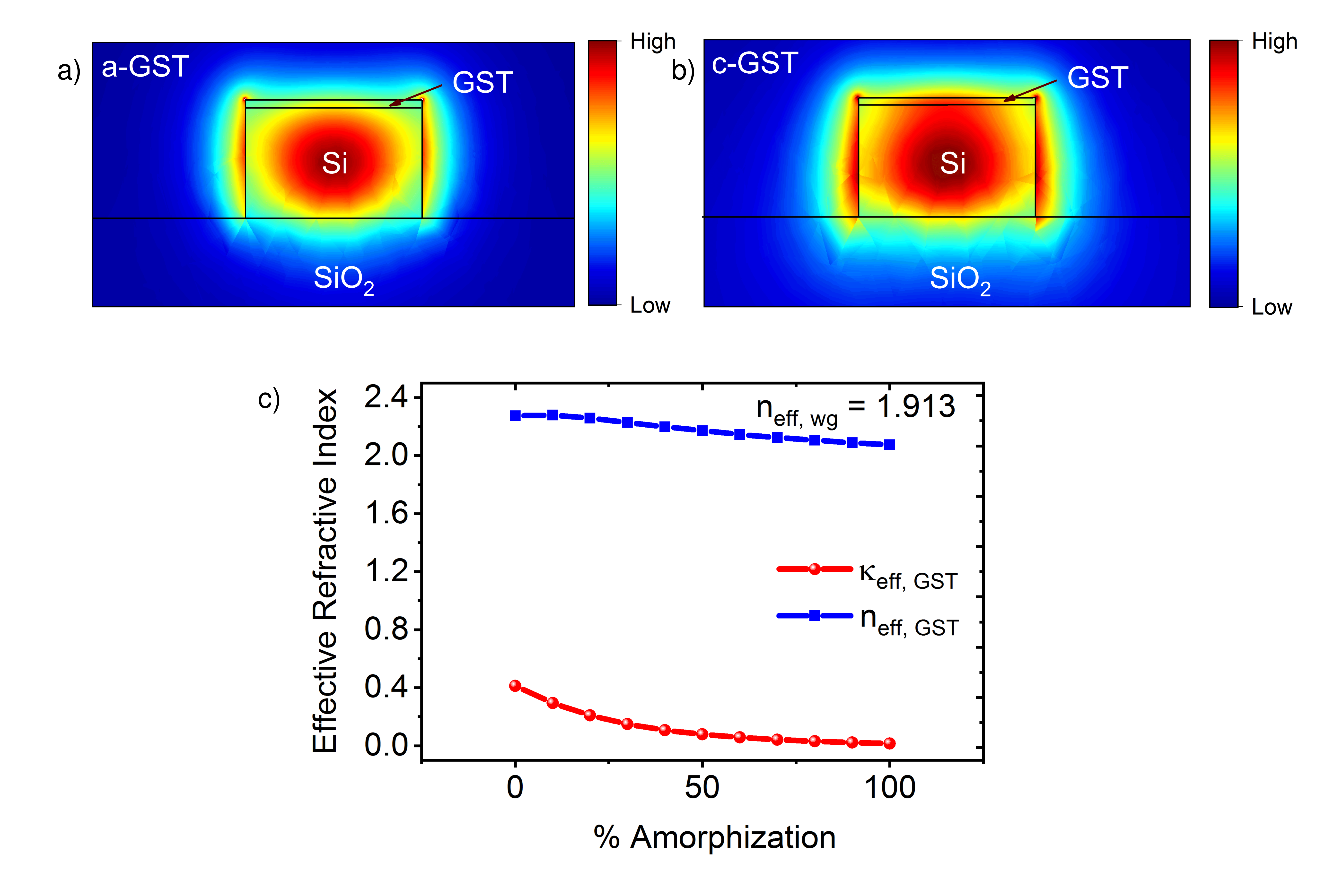}
		\vspace{-7mm}
		\caption{Cross-section view of Fundamental Mode profiles for a GST-embedded Si-SiO\textsubscript{2} waveguide section for (a) a-GST and (b) c-GST showing visible contrast in optical absorption for the two boundary states of GST. (c) The variation of the real ($n_{eff,GST})$) and imaginary ($\kappa_{eff,GST}$) refractive indices of GST with degree of crystallization.}
		\label{fig:mode}
	\end{figure}
\begin{equation}
\frac{\epsilon_{eff}(p)-1}{\epsilon_{eff}(p)+2}=p\times\frac{\epsilon_{c}-1}{\epsilon_{c}+2}+(1-p)\times\frac{\epsilon_{a}-1}{\epsilon_{a}+2}
\end{equation}
where $\epsilon_{c}$ and $\epsilon_{a}$ are the complex permittivites of c-GST and a-GST respectively calculated from the refractive indices of GST\cite{pernice2012photonic} by $\sqrt{\epsilon(\lambda)}=n+i\kappa$. $p$ is the degree of crystallization. Thus, the different levels of crystallization of GST leads to various levels of $\kappa_{eff,GST}$ thus leading to different levels of transmission. We leverage the multi-level transmission to implement an all-photonic synapse. Considering an incident optical pulse of power $P_{in}$, the synaptic functionality is realized such that the output power $P_{out}$ is given by:
\begin{equation}
P_{out} = T_{\lambda_m}P_{in}
\end{equation}

where $T_{\lambda_m}$ is the transmission at resonant wavelength $\lambda_m$. $T_{\lambda_m}$ represents the weight of the synapse and the various levels of transmission with varying degree of crystallization states of GST can be leveraged to represent a entire range of synaptic weights with appropriate discretization. We critically couple the resonator to the amorphous state such that the transmission is minimum in the amorphous state and increases with the degree of crystallization. While individual synapses represent a simple multiplication, the weighted inputs from multiple synapses are received by a neuron as shown in Fig. \ref{fig:Syndev} (a). To emulate such a behavior, it is important to connect these synapses in an integrated fashion. Such a synaptic network would perform the most ubiquitous functionality of any neural network, a dot-product.

\section{Photonic Dot Product Engine}
We leverage the characteristics of the proposed non-volatile photonic synaptic device to map the synaptic weights of a neural network in a Photonic Synaptic Network capable of performing the dot-product of the inputs and the weights. 
\begin{figure}[t]
		\centering
		\includegraphics[width=3.6in, keepaspectratio]{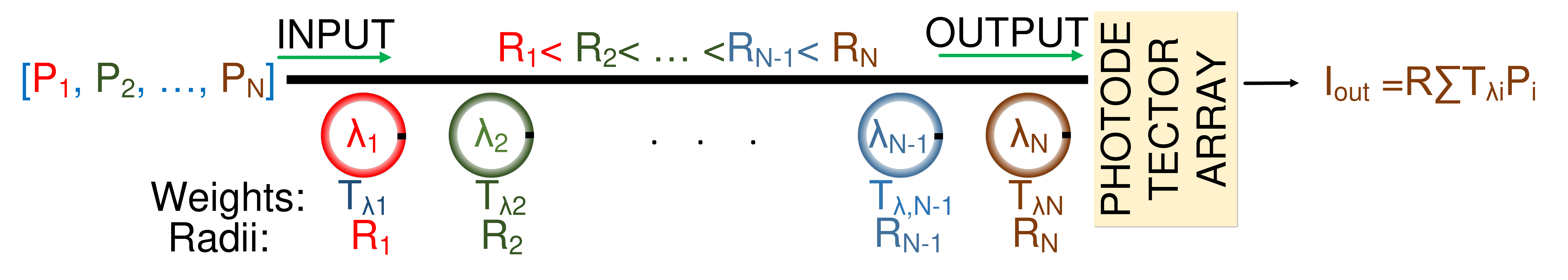}
		\caption{Synaptic dot product engine showing arrangement of ring resonators with increasing radii representing the transmission vector $T_\lambda = \{T_{\lambda_1}, \dots, T_{\lambda_N}\}$. WDM signals gets modulated by weights corresponding to respective wavelength and the photodetector array collects the signals to generate a current $I_{out}$ representing the dot product of transmission vector $T_\lambda$ and inputs $P = \{P_1,\dots, P_N\}$.}
		\label{fig:SA1row}
	\end{figure}
\subsection{Network Design} 
We leverage the Wavelength Division Multiplexing (WDM) technique to compute dot product operations between incoming spikes and synaptic weights. We represent the synaptic weights in terms of the transmission $T_\lambda$ of the microring resonator as discussed in the previous section. To represent multiple wavelengths, we use multiple ring resonators of increasing ring radii to represent different synapses in a row as shown in Fig. \ref{fig:SA1row}. The number of synapses ($N$) in each row is dependent on the Free Spectral Range (FSR) of the ring resonator and this governs the dimension of the input vector of the dot product engine. A WDM spike enters the straight waveguide through the `INPUT' port and the GST element on each ring resonator modulates the amplitude of corresponding wavelength by the representative synaptic weight according to Eqn. (5). Thus at the `OUTPUT' port we obtain a multi-wavelength spike comprising of different $T_{\lambda_i}P_i$ products corresponding to different wavelengths. This spike is then fed to a photodiode array (PD) which produces a current given by the sum of all the amplitudes given by:
\begin{equation}
I_{out} = R \sum_i{T_{\lambda_i}P_i}
\end{equation}
where R is the responsivity of the PD expressed as A/W. This current is equal to the dot product of the input vector $P$ and weight vector $T_\lambda$. The operation is illustrated in Fig. \ref{fig:SA1row}. 

\subsection{Synapse Design constraints}
Using the WDM technique for the proposed photonic synaptic array imposes certain constraints on the design of the synaptic devices. For accurate dot-product operation, it is necessary to achieve significant isolation between the channels in order to minimize channel-to-channel interaction. The important parameters which constrain the design space of the synaptic device are finesse (F) and channel spacing ($\lambda_{diff}$). Finesse is the ratio of free spectral range (FSR) and full-width at half maximum (FWHM). For a single bus ring resonator, FWHM and FSR are expressed as \cite{bogaerts2012silicon}:
\begin{gather}
FWHM = \frac{(1-ra)\lambda_m^2}{\pi n_gL\sqrt{ra}} \\
FSR = \frac{\lambda_m^2}{n_gL} \\
Finesse = \frac{FSR}{FWHM}
\end{gather}
where $L = 2\pi R_{ring}$ is the circumference of the ring, $n_g$ is the group index and rest of the parameters bear the same meaning as defined earlier. The interference due to adjacent channels can be modeled as:
\begin{equation}
\begin{aligned}
T'_{\lambda_i}|_{\lambda=\lambda_i} &= T_{\lambda_i}|_{\lambda=\lambda_i}\times T_{\lambda_{i}}|_{\lambda=\lambda_{i+1}}\times T_{\lambda_{i}}|_{\lambda=\lambda_{i-1}} \\
T'_{\lambda_i}|_{\lambda=\lambda_i} &= \alpha_{\lambda_i}T_{\lambda_i}|_{\lambda=\lambda_i}
\end{aligned}
\end{equation}
Here, $T'_{\lambda_i}|_{\lambda=\lambda_i}$ is the modified transmission due to interference from the adjacent resonant wavelengths, $T_{\lambda_i}|_{{\lambda=\lambda_i, \lambda_{i+1}, \lambda=\lambda_{i-1}}}$  are the transmissions of $i^{th}$ ring at the $i^{th}$, ${(i+1)}^{th}$ and ${(i-1)}^{th}$ resonant wavelengths respectively. $\alpha_{\lambda_i}$ represents the non-ideal factor which should ideally be close to 1. $\alpha_{\lambda_i}$ decreases with decreasing channel spacing ($\lambda_{diff}$) and increasing FWHM. For our design, we decided the minimum radius of the ring to be 1.5 $\mu m$ in order to achieve a high density synaptic array for better scalability. Rings of similar size have been demonstrated previously \cite{Xu_2008} with certain modifications that we will discuss next. The rest of the parameters concerning the synapses were chosen to maximize the number of rings in a single row (N) while maintaining $\alpha_{\lambda_i}$ close to 1 under the condition that $N\lambda_{diff} <FSR$. 

A number of challenges arise for rings of radius comparable to the wavelength of operation. Firstly, to achieve a critical coupling in the low-loss amorphous state, the power coupling gap between the bus and the ring waveguide needs to be small ($<100 nm$). This is because the interaction length between the ring and the straight waveguide is quite short and hence to achieve reasonable coupling, even to match the small intrinsic loss in the ring in low-loss amorphous state of GST, we require a small power coupling gap. Such gaps become extremely difficult to fabricate. An alternative to using lower gaps has been demonstrated \citep{Xu_2008} for rings of small radii. Reducing the width of the bus waveguide increases the spatial period of the propagating mode due to the lower effective refractive index. This results in a better phase match with the mode in the tightly curved ring waveguide. For the rest of our analysis, we have used a bus waveguide of width 0.35 $\mu m$ and a coupling gap of 135 $nm$. 
\section{Photonic Integrate-and-Fire Neurons}
\begin{figure}[t]
		\centering
		\includegraphics[width=3.6in, keepaspectratio]{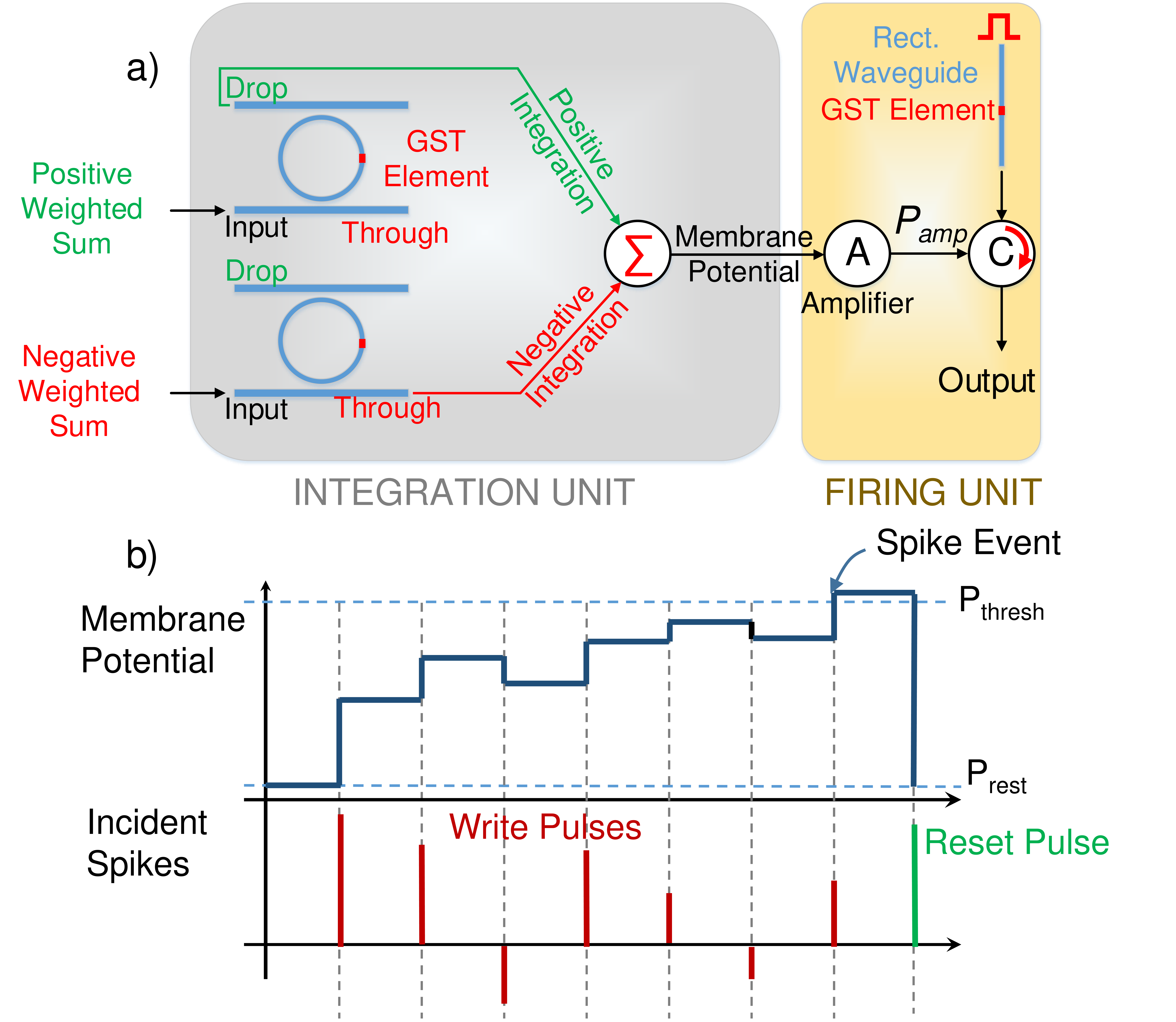}
		\caption{(a) Schematic of a bipolar integrate and fire neuron based on GST-Embedded Ring resonator devices showing the integration and firing unit. (b) Timing diagram showing the integration of membrane potential for various incident pulses
demonstrating the operation of the proposed neuron}
		\label{fig:neuron}
	\end{figure}
The proposed photonic dot-product engine needs to be interfaced with spiking neurons to realize a Photonic SNN inferencing platform. In this work, we explore a Photonic `Integrate-and-Fire' neuron that we have proposed previously \cite{chakraborty2018all}.  
We revisit the concept of a Photonic Integrate-and-Fire Neuron explored in our previous work \cite{chakraborty2018all}. The neuron consists of an `Integration Unit' and a `Firing Unit'. The `Integration unit' of the neuron consists of two add-drop ring resonators with GST deposited on top of each as shown in Fig. \ref{fig:neuron} (a). The purpose of the two ring resonators is to perform bipolar integration, i.e., the respective devices are fed by positive and negative weighted sums from the synapses to perform integration in the appropriate direction. The significance of positive and negative weighted sums would be clearer in the next section. The neuron operates in alternate `write' and `read' cycles. The GST elements on the ring resonators are initially in crystalline state. With incident `write' pulses, the GST element begins to get partially amorphized. During the `read' phase, with partial amorphization, transmission at the `THROUGH' port of each ring resonator decreases and that at the `DROP' port increases. Essentially, with incoming pulses, the transmission through the `DROP' and `THROUGH' ports get positively and negatively integrated respectively. These properties of the device can be combined to mimic the behavior of a bipolar integrate and fire neuron. The `DROP' and `THROUGH' port of the positive and negative integrating ring resonator respectively are connected to an inteferometer. The output of the interferometer represents the membrane potential of the spiking neuron. To perform the thresholding action, the membrane potential is fed to the `Firing unit' of the neuron. This unit consists of an amplifier, a circulator and a rectangular waveguide with GST deposited on top. During the `read' phase of the neuron, the resulting membrane potential after being amplified and directed by the circulator towards the rectangular waveguide, attempts to amorphize the initially crystalline GST element on the rectangular waveguide. Initially, the output of the amplifier A ($P_{amp}$) is insufficient to amorphize the GST on rectangular waveguide and hence rendering it unable to transmit an output spike. However, when the membrane potential integrates enough to the cross the threshold, on incidence of several ‘write’ pulses, $P_{amp}$ is ensured to be high enough to amorphize the GST on the rectangular waveguide, thus enabling it to transmit a spike. Once the neuron fires, a `RESET' pulse resets the states of the devices to their initial states and the membrane potential drops to the resting potential ($P_{rest}$) as shown in Fig. \ref{fig:neuron} (b). Further details of the writing and reading schemes have been presented in \cite{chakraborty2018all}.

\section{Operation of All-Photonic Spiking Neural Network}
\begin{figure}[t]
		\centering
		\includegraphics[width=3.5in, keepaspectratio]{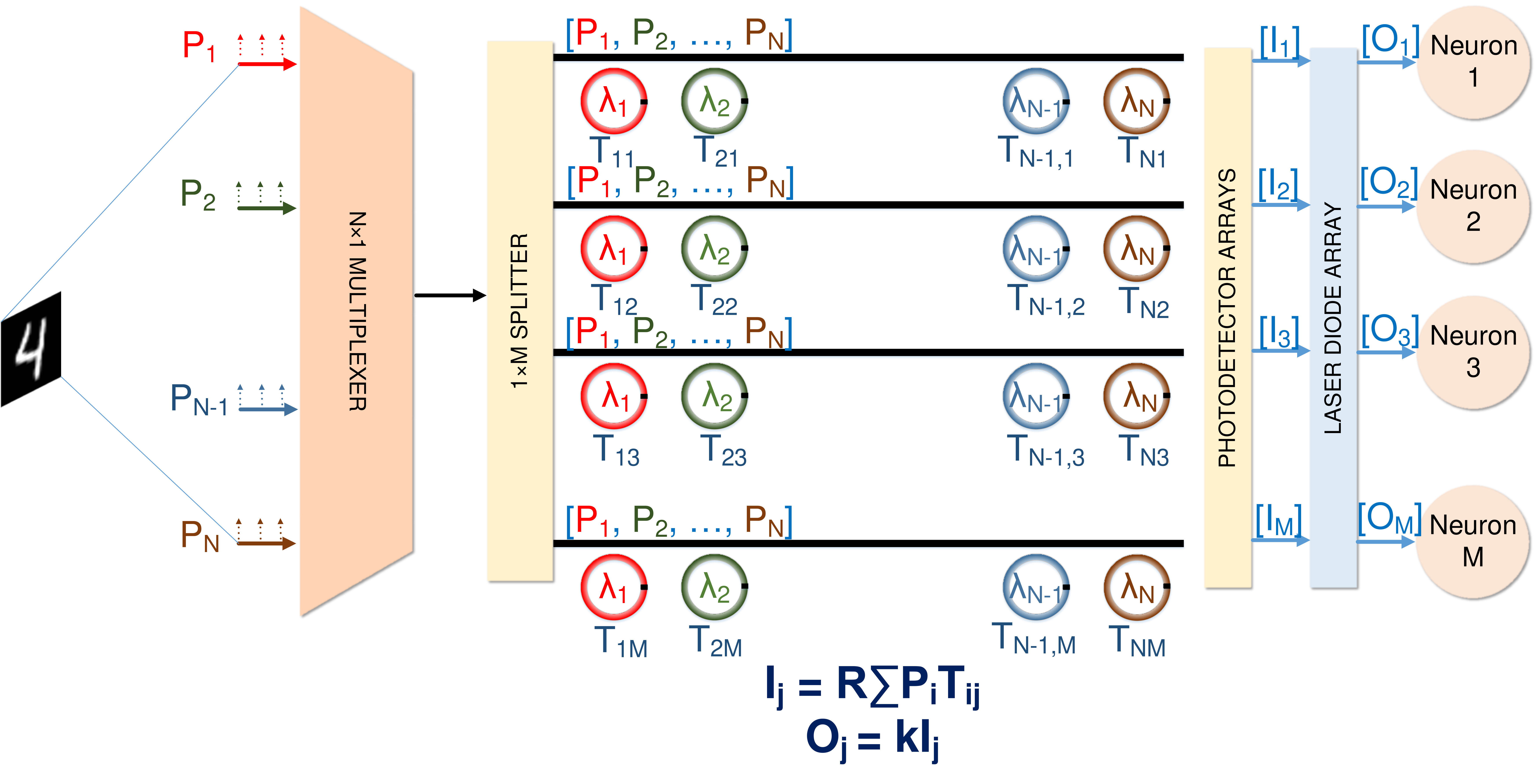}
		\caption{Synaptic dot product engine showing arrangement of ring resonators with increasing radii representing the transmission vector $T_\lambda = \{T_{\lambda_1}, \dots, T_{\lambda_N}\}$. WDM signals gets modulated by weights corresponding to respective wavelength and the photodetector array collects the signals to generate a current $I_{out}$ representing the dot product of transmission vector $T_\lambda$ and inputs $P = \{P_1,\dots, P_N\}$. $k$ is an amplification factor.}
		\label{fig:SAfull}
	\end{figure}

Implementation of a SNN based on the Photonic Dot-Product Engine (PDPE) and `integrate-and-fire' neurons described above involves integration of the proposed structures. As elucidated above, the basic computational function of a neural network is a dot product. To realize parallel instances of such a functionality using the aforementioned PDPE, we use a splitter (SPL) to feed the WDM input spikes to multiple PDPE rows with the input vector and obtain the dot-products of each rows from respective PD arrays as shown in Fig. \ref{fig:SAfull}. Essentially, the output vector thus obtained from the PD arrays gives us the multiplication of the vector of input spikes $P_i$ with a $N\times M$ synaptic network $T_{ij}$. 
The M outputs $I_j$ obtained from the PD arrays are fed to laser diodes (LD) which converts the electrical current to optical spikes thus completing the parallel dot-product operations and can be represented as:
\begin{equation}
\begin{bmatrix}O_1\\O_2\\\vdots\\O_M\end{bmatrix} \propto \begin{bmatrix}P_1&P_2&\dots &P_N\end{bmatrix}\begin{bmatrix}T_{11} & T_{12} & \dots & T_{1M}\\T_{21} & T_{22}& \dots & T_{2M}\\\vdots & \vdots &\ddots &\vdots\\T_{N1} & T_{N2} & \dots & T_{NM}\end{bmatrix}
\end{equation}
\begin{figure*}[t]
		\centering
		\includegraphics[width=\textwidth, keepaspectratio]{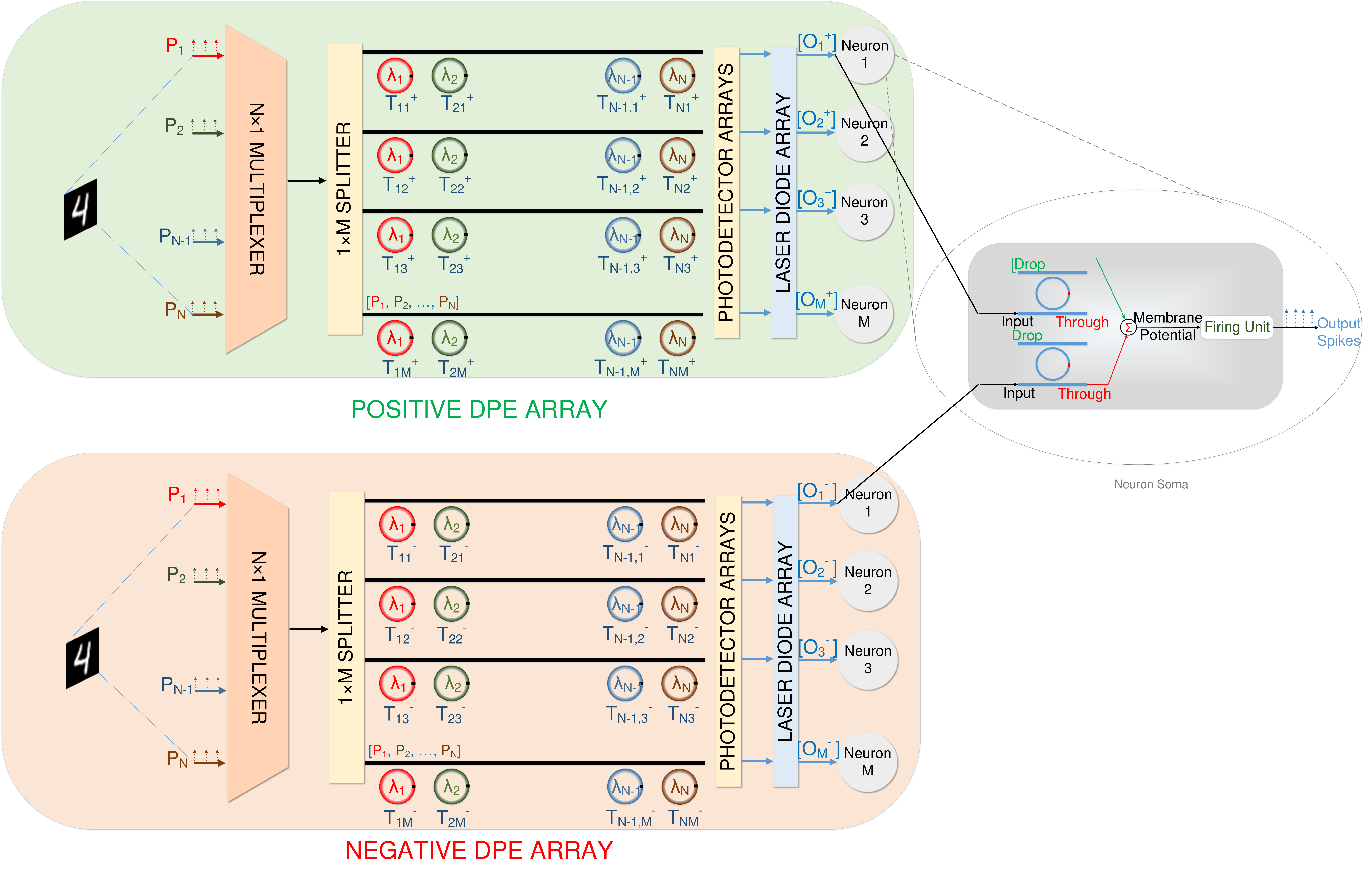}
		\caption{Schematic of an All-Photonic Spiking Neural Network. Two DPE arrays are deployed to represent the positive and negative components of the weights. The outputs of the DPE arrays are converted to optical spikes and passed to integrate-and-fire neurons. The structure of an integrate-and-fire neuron is illustrated in a circle. Each neuron has two inputs corresponding outputs from the positive and negative DPE arrays. The neuron outputs a spike when the membrane potential crosses its threshold.}
		\label{fig:SNNfull}
\end{figure*}
We now present how such a photonic synaptic network based can be integrated with the proposed bipolar IF Neurons to realize a photonic SNN. The schematic of such a photonic SNN is illustrated in Fig. \ref{fig:SNNfull}. To account for negative weights in a neural network, we represent the element of the weight matrix $T$ to be comprised of a positive and negative component: 
\begin{equation}
\begin{aligned}
T_{ij} &= T_{ij}^+ + T_{ij}^- \\
T_{ij}^+ &= T_{ij}, T_{ij}^- = T_{low},  \textnormal{when } T_{ij}>0\\
T_{ij}^+ &= T_{low}, T_{ij}^- = |T_{ij}|, \textnormal{when } T_{ij}<0\\
\end{aligned}
\end{equation}
Here $T_{low}$ is the transmission corresponding to the lowest programmable state considered. Two PDPE arrays are deployed for mapping the positive and negative components respectively as depicted in Fig. \ref{fig:SNNfull}. The dot-product outputs from the LD arrays of the two DPE arrays can be represented as:
\begin{equation}
\begin{aligned}
O_{j}^+ &= \sum_iP_iT_{ij}^+ \\
O_{j}^- &= \sum_iP_iT_{ij}^- \\
\end{aligned}
\end{equation}
These outputs from the $j^{th}$ rows are received by the $j^{th}$ IF neuron discussed earlier. The outputs from the positive and negative PDPE arrays are received by the positive and negative integrating ring resonators in the neuron respectively. The two ring resonators integrate in the opposite direction based on the two inputs and the resulting integration mimics the desired integration that a biological `integrate-and-fire' neuron performs, given by:
\begin{equation}
V_{mem,j}[t] = V_{mem,j}[t-1]+\sum_iP_iT_{ij}
\end{equation}
\begin{table}[t]
\centering
\caption{Simulation Parameters}
\label{param}
\begin{tabularx}{\columnwidth}{|X|X|}

\hline
Parameters                                                           & Values                       \\ \hline\hline
Si Ring Waveguide X-Section                                           & 0.45$\times$0.25 ${\mu m}^2$ \\
Si Bus Waveguide X-Section                                           & 0.35$\times$0.25 ${\mu m}^2$ \\
Coupling Gap ($L_{gap}$)                                             & 0.135 $\mu m$                \\
GST Length ($L_{GST}$)                                               & 170 nm - 220nm               \\
GST Thickness ($t_{GST}$)                                            & 10 nm                        \\
GST Width ($W_{GST}$)                                                & 0.44 $\mu  m$                \\
Si Refractive Index ($n_{Si}$) \cite{aspnes1983dielectric}                                        & 3.5                          \\
SiO\textsubscript{2} Refractive Index ($n_{SiO_2}$) \cite{malitson1965interspecimen}   & 1.4                          \\
c-GST Refractive Index ($n_{c-GST} +i\kappa_{c-GST}$) \cite{kim1998variation}                                           & 7.2+1.9i                     \\
a-GST Refractive Index ($n_{a-GST}+i\kappa_{a-GST}$) \cite{kim1998variation}                                           & 4.6+0.18i                    \\ \hline
\end{tabularx}
\end{table}
Here, $\sum_iP_iT_{ij} = \sum_i(P_iT_{ij}^+-P_iT_{ij}^-)$. $V_{mem,j}[t]$ is the internal state or the membrane potential of the $j^{th}$ neuron at time $t$. The resulting membrane potential is passed to a Firing Unit as described in Fig. \ref{fig:neuron} such that the neuron produces an output spike once the $V_{mem,j}[t]$ reaches a threshold. The output spikes from all the neurons of the current layer are then fed to the next synaptic array layer. Fig. \ref{fig:SNNfull} delineates the operation of basic building blocks of a neural network. We perform large scale system-level simulations by emulating the behavorial model of the proposed spike processing system to assess the performance of neuromorphic systems based on this fabric. \par
It is important to consider the architecture-level facets of any computing primitive. The proposed design is analogous to memristive crossbars, where the high fan-in into the neurons is resolved by the inherent parallelism of the computing framework. In our design, each neuron receives two inputs, from the positive and negative synaptic array, and the output of that neuron is fed to one of the 16 inputs of the synaptic array of the next layer. In reality, neural networks are of far bigger sizes than what the proposed design can accommodate. As a result, multiple instances of the proposed primitive can be used with time-multiplexing to perform the entire vector-matrix multiplication operation. The partial sums from these instances are collected and added before being fed to the neuron. Output from a neuron is again served as inputs to the synaptic arrays storing the weights of the next layer of the neural network. Similar architectures have been explored using memristive technologies \mbox{\cite{shafiee2016isaac, ankit2017resparc}}. This work is concerned with device and circuit primitive of a spike-based photonic non-volatile inferencing engine which will act as a computing core of a large-scale system similar to technologies in the electrical domain.
\section{Results}
\subsection{Simulation Framework}
\subsubsection{Device Simulations}
We evaluated the performance of the proposed all-photonic SNN fabric by designing a device-circuit-algorithm co-simulation framework. First, the device characteristics of each ring resonator in a DPE row is simulated for 4 different degrees of crystallization of the GST element using commercial-grade simulator Lumerical FDTD Solutions\cite{lumerical} based on the finite-difference time-domain (FDTD) method. The fixed parameters used for these simulations are listed in Table \ref{param}. The mode-profiles were obtained through Electromagnetic simulations using the Finite Element method in COMSOL Multiphysics \cite{comsol}. 

\subsubsection{Device to System Framework}
The device characteristics, obtained from the FDTD simulations are analyzed and a Gaussian fit is applied on the data for interpolation. We develop a device to system co-design framework by building behavorial models of the proposed synapses and neurons based on the fitted device characteristics. The models are used to evaluate the inferencing performance of the standard neural network topology on standard digit recognition task based on the MNIST dataset using the Deep Learning Toolbox\cite{palm2012prediction} in MATLAB. The MNIST dataset consists of 60000 images in the training set and 10000 images in the testing set.  
\subsection{Device Simulations}
\begin{figure}[t]
		\centering
		\includegraphics[width=3.6in, keepaspectratio]{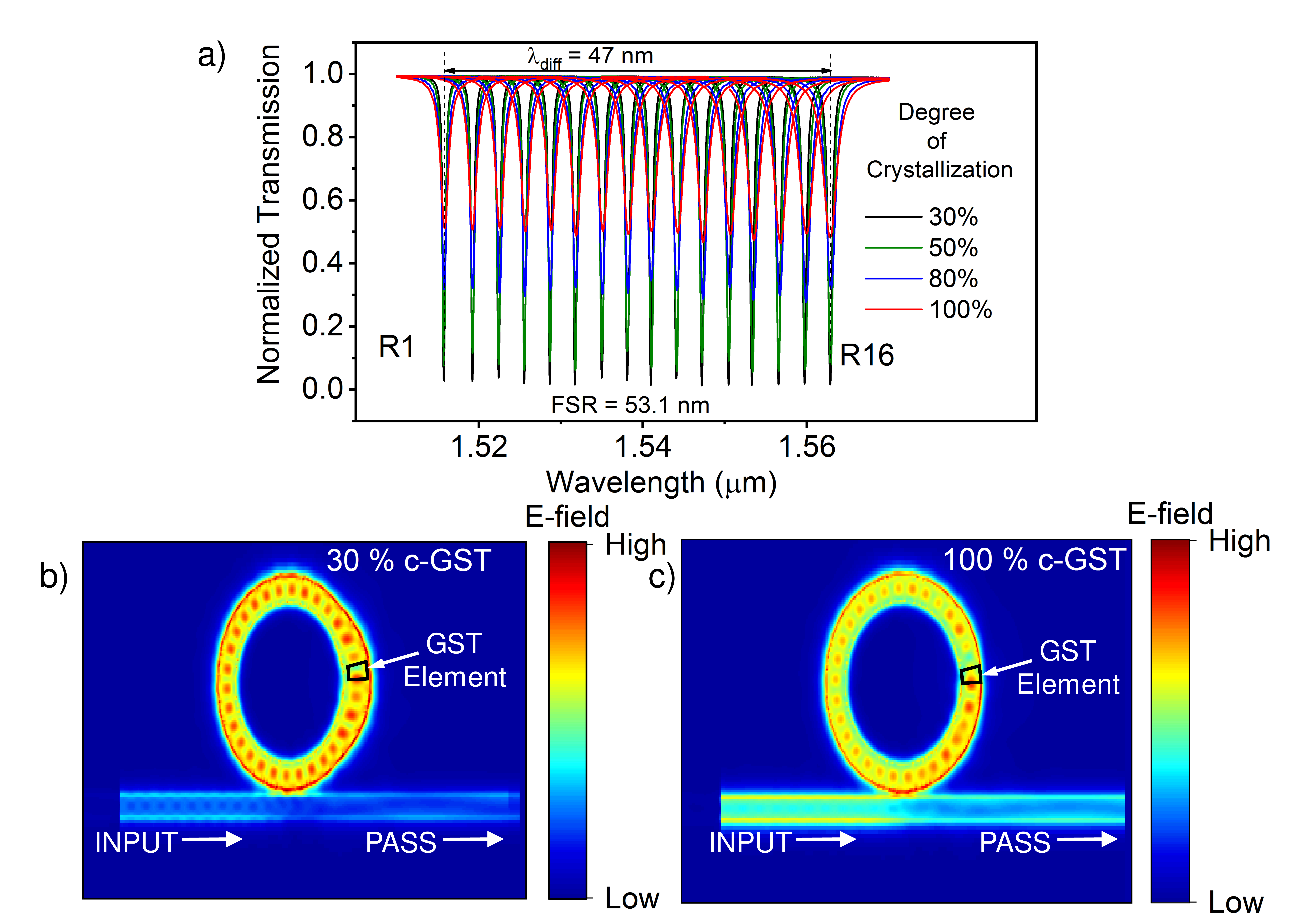}
		\caption{(a) Normalized transmission for 16 different rings for 4 degrees of crystallization (30 \%, 50 \%, 80 \%, 100 \%) showing a decreasing trend with decreasing degree of crystallization. The range of wavelength for the 16 rings is less than the FSR for the design. (b) and (c) shows the electric field profile in the ring resonator system showing visible contrast in optical absorption and field transmission at the `PASS' port in the GST element for c-GST and 30\% c-GST respectively.}
		\label{fig:transmission}
	\end{figure}
We considered 16 ring resonators of radii linearly increasing from 1.5 $\mu m$ to 1.59 $\mu m$ in any particular DPE row. The choice of number of devices, $N$, in a single row is discussed earlier. The length of the GST element is increased accordingly and chosen iteratively to ensure uniform transmission characteristics across the wavelength range of operation. We performed FDTD simulations for each device with 4 different degrees of crystallization of GST (30\%, 50\%, 80\%, 100\%) and the observed transmission characteristics for the rings are shown in Fig. \ref{fig:transmission} (a). Expectedly, the transmission for each device decreases with decreasing degree of crystallization. The observed FSR was 53.1 nm and difference between the highest and lowest resonant wavelength was 47nm, which is well within the FSR, thus ensuring no interference from resonant wavelengths beyond the region of operation. Fig. \ref{fig:transmission} (b) and (c) show the contrast in electric field absorption by the GST element in the ring resonator for 30\% and 100\% crystallized GST. We observe certain variations across different wavelengths which can be minimized by further adjustments of lengths of the GST element. However, from the perspective of neuromorphic applications, these variations prove to be insignificant. We will explore the impact of such variations in our evaluation of the proposed neuromorphic processing engine.  
\begin{figure}[t]
		\centering
		\includegraphics[width=2.7in, keepaspectratio]{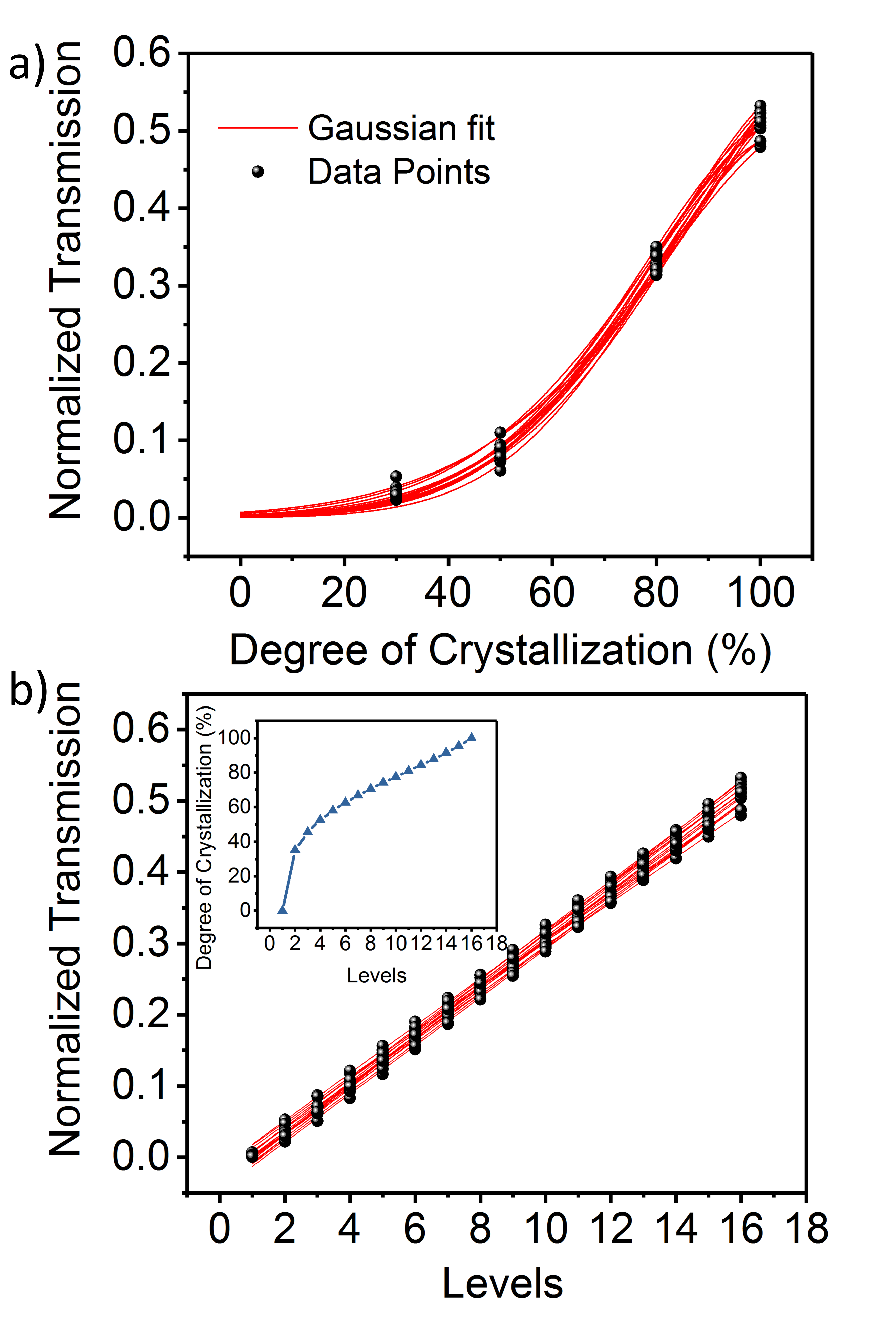}
		\caption{(a) Gaussian fit of simulated data points across degrees of crystallization ranging from 0 \% and 100 \%. (b) Linearly varying transmission across 16 different programmable states (Levels) of the GST. Inset shows the degrees of crystallization corresponding to the Levels.}
		\label{fig:levels}
	\end{figure}
We exploit the dependence of transmission on degree of crystallization to realize the synaptic behavior of the rings. Fig. \ref{fig:levels} (a) shows the Gaussian fit of the simulated data across degrees of crystallization varying from 0\% to 100\%. Note, the Gaussian fit provides a fairly accurate representation of the observed data and is a powerful tool to speed up our analysis in light of the computationally expensive FDTD simulations. It can be observed that transmission has a non-linear relationship with $p$ and hence, operation of the rings as synapses would require the GST element to be programmed to states with non-linearly increasing $p$. This can be achieved with appropriate amplitude of the programming stimulus. Fig. \ref{fig:levels} (b) shows the transmission levels for each ring corresponding to 16 discretized programmable states or Levels. The degrees of crystallization, $p$, for each state is shown in the inset of Fig. \ref{fig:levels} (b). The linear relationship between transmission and Levels is a necessity for the target application, i.e., a dot-product operation for neuromorphic computing which led us to the choice of programmable states with the non-linear distribution of $p$. 

\subsection{Interference Errors}
\begin{figure}[t]
		\centering
		\includegraphics[width=3.3in, keepaspectratio]{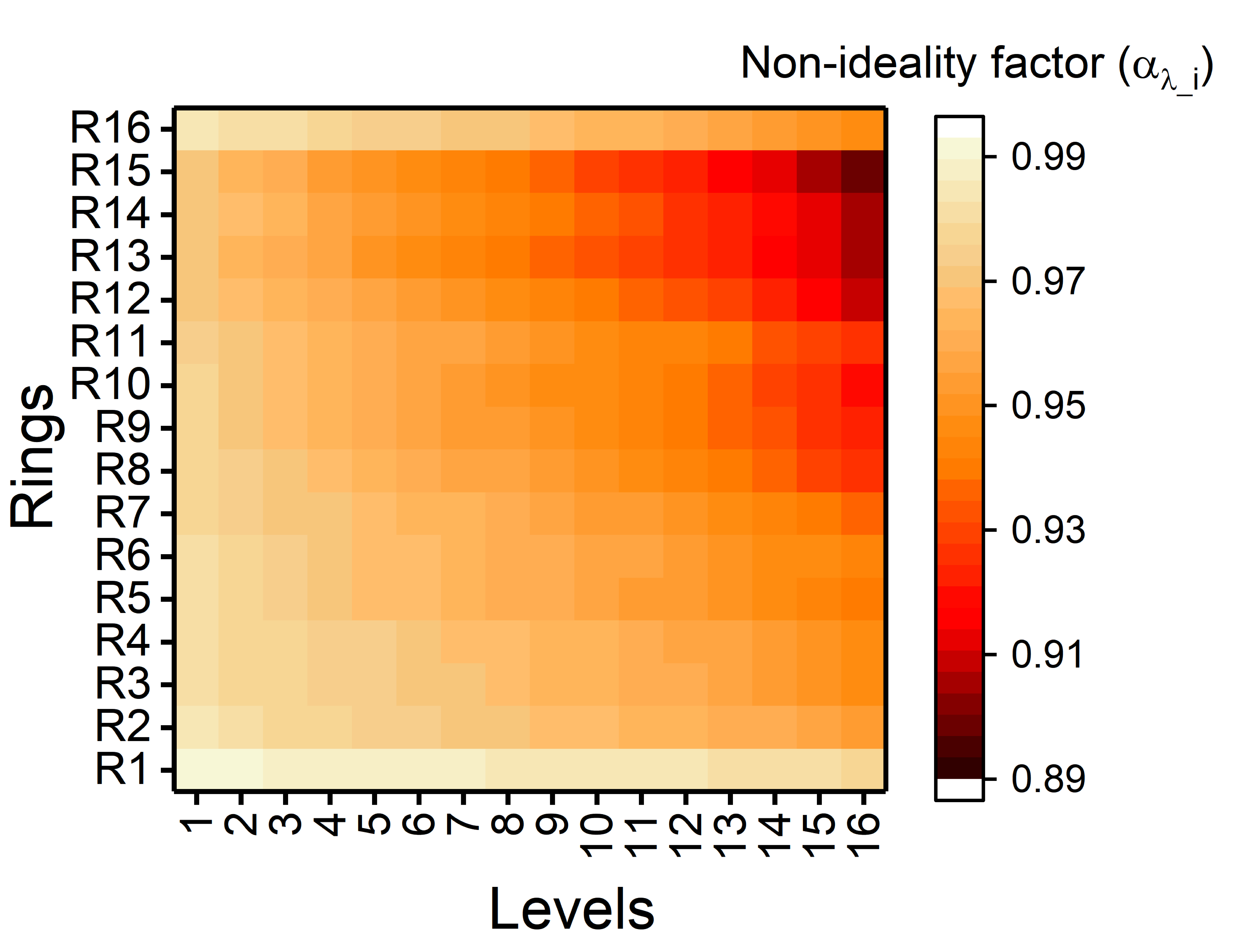}
		\caption{Map of non-ideality factor ($\alpha_{\lambda_i}$) arising due to interference from adjacent rings for each ring in the DPE row.}
		\label{fig:errors}
	\end{figure}

The transmission characteristics of the different rings for varying states of the GST element is used to evaluate the accuracy of the dot-product operation performed using the proposed synaptic network.  The error in the computation stems from the premise of overlapping frequency response between adjacent channels. The advantage of the proposed implementation over electrical counterparts is that in the electrical domain, the losses due to line resistance is a function of input and the weights thus rendering them difficult to model. The impact of the error in this setup is only dependent on the weight level and hence, can be easily modeled, analyzed and even corrected in light of the proposed application. In Eqn. 9, we have formulated a behavorial model of the error arising from interference due to adjacent channels. Fig. \ref{fig:errors} shows the map of non-ideality factor $\alpha_{\lambda_i}$ for all 16 rings for 16 different levels. This was calculated through fitting of the extracted $\alpha_{\lambda_i}$ from Fig. \ref{fig:transmission} (a) based on Eqn. 9. We observe that errors are highest for rings of higher radius and for the highest levels. This can be attributed to higher FWHM for rings of higher radius due to the longer lengths of the GST element used to achieve uniform transmission levels across the operating range of wavelength. We include these error characteristics corresponding to each ring for our system level evaluation of the proposed photonic SNN inferencing framework.
\subsection{System Level SNN performance}
\begin{figure}[t]
		\centering
		\includegraphics[width=2.8in, keepaspectratio]{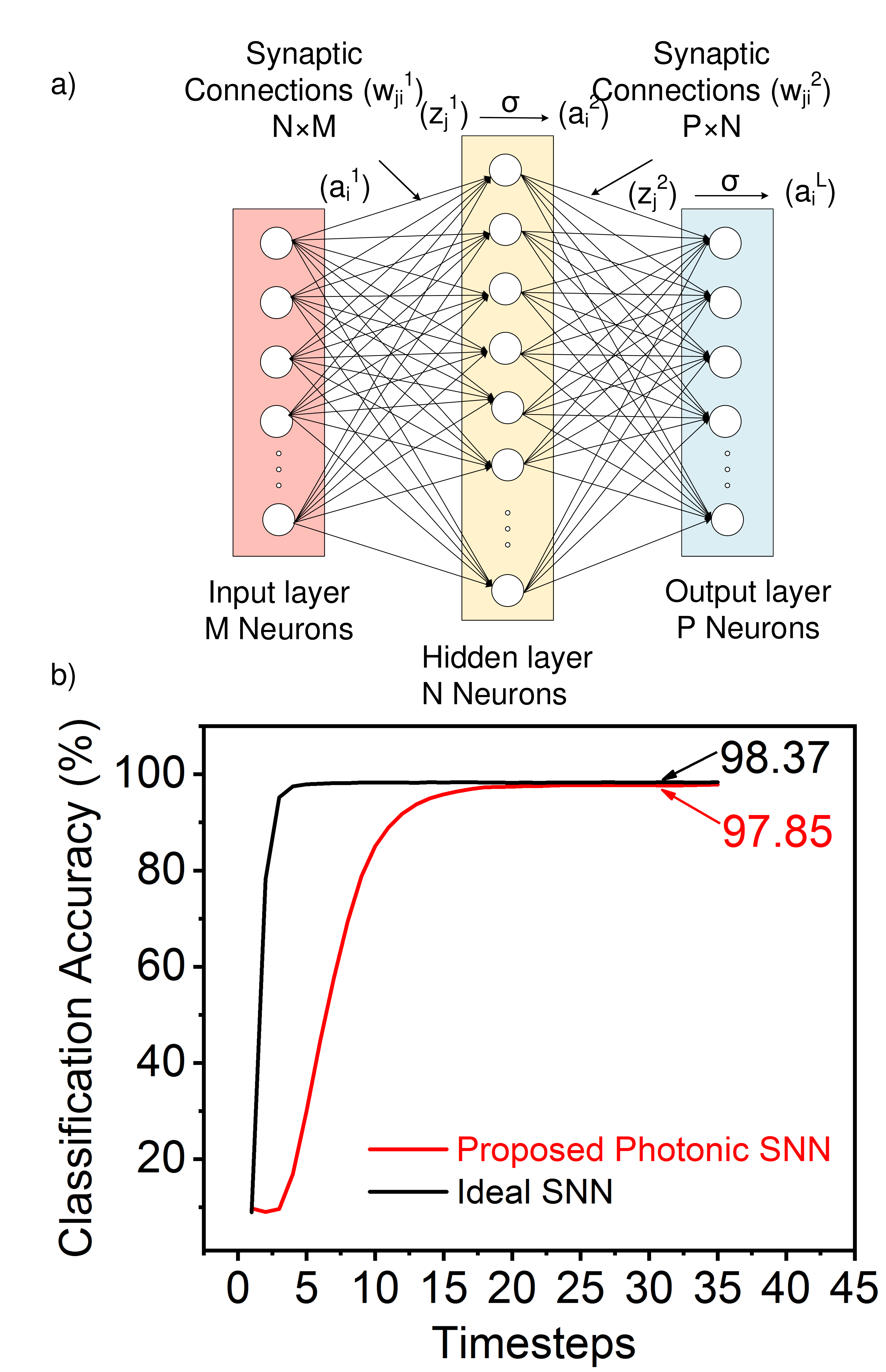}
		\caption{(a) Fully connected neural network topology consisting of an input layer (M), a hidden layer (N) and an output layer (P) of neurons. The resulting synaptic networks are of sizes $N\times M$ and $P\times N$(b) Evolution of classification accuracy of handwritten digit recognition task based of MNIST dataset comparing our proposed Photonic SNN to ideal SNN performance. Here ideal SNN corresponds to software-level functionalities without considering device characteristics.}
		\label{fig:SNNacc}
	\end{figure}
We develop a device to algorithm level framework to perform system level analysis of the photonic SNN implementation. A SNN, like any other neural network, consists of multiple layers of neurons connected through synapses. The unique property of SNNs is that the inputs to the network are discretized spike events instead of analog values. The synapses act as weights which get multiplied with amplitude of the incoming stimulus and the resulting weighted-sum, i.e., dot-product of all impulses coming from different synapses is received by the neuron. We map the device characteristics of each individual synapse and `integrate-and-fire' spiking neurons discussed previously to explore the validity of operation of the proposed devices as synapses and neurons in such a SNN. Let us now explain how we perform the evaluation of a SNN on the proposed PCM-based photonic inferencing framework. We consider a fully connected neural network consisting of 3 layers, namely, the input layer, the hidden layer and output layer as shown in Fig. \ref{fig:SNNacc} (a). This type of topology is well explored \cite{diehl2015fast}. For our analysis, we consider a network with $M = 784$, $N=500$, $P=10$. We analyze the accuracy of such a network in a standard handwritten digit recognition task based on the MNIST dataset \cite{lecun-mnisthandwrittendigit-2010}. A popular way of implementing spike-based inferencing systems is to train a network as an Artificial Neural Network  (ANN)  and  then  convert  it  to  a  SNN  by  well  explored  conversion  algorithms \mbox{\cite{diehl2015fast, sengupta2018going}}.  The weights of the network are trained using the Backpropagation algorithm \mbox{\cite{rumelhart1986learning}} as in case of Artificial Neural Networks (ANN). The neurons in ANNs are usually non-linear mathematical functions, such as Rectified Linear Units (ReLU) \mbox{\cite{nair2010rectified}}, sigmoid  or  tanh  with  ReLU  being  the  most  popularly  chosen  neuron  functionality.  During  conversion,  an  artificial neuron  with ReLU  functionality  can  be  directly  converted  to  an  IF  neuron,  mathematically \mbox{\cite{diehl2015fast}}. The details of the operation of the IF neuron has been elucidated in our earlier work\cite{chakraborty2018all}. The trained weights of the network after the ANN is converted to a SNN are mapped to the observed characteristics of each synaptic device in the proposed synaptic network. The synaptic network has the provision of operating 16 synapses simultaneously. To perform the dot-product of larger dimensions, the synaptic network needs to be time-multiplexed as discussed earlier. To simulate large-dimension operations with the proposed synaptic network, we repeat the device characteristics every 16 synapses. The weights of the network can be negative. To account for negative weights, two dot-product engines are deployed, shown in Fig. \ref{fig:SNNfull} as described earlier. 

The pixels of input images of size $28\times28$ are divided into streams of spikes whose frequency is proportional to the pixel intensity. At every time-step, the input can either be `0' when there is no spike or `1' in the event of a spike. The behavorial model of the SNN inferencing framework described above was implemented using the MATLAB Deep Learning Toolbox \cite{palm2012prediction} using the network topology shown in Fig. \ref{fig:SNNacc} (a). The network is evaluated at every time-step by passing the inputs through the forward path from the input layer to the output layer through the synaptic network and activity of the network was recorded. Finally, the output neuron with the highest spiking activity is compared with the label of the input image to determine the accuracy of the recognition system. The classification performance of the proposed photonic SNN is compared with an ideal SNN in Fig. \ref{fig:SNNacc} (b). Here, ideal SNN essentially means software-level evaluation without taking device characteristics into consideration.  We observe that there is a degradation in accuracy of 0.52 \% after 35 time-steps from the ideal case arising from the different variations in device characteristics discussed earlier. To note, the concept of time-steps here correspond to how many times we evaluate the network over the Poisson-distributed input spikes generated from the image. The duration of a time-step is not relevant in this context as we do not include any temporal dynamics in the system. We further attempted to isolate the contribution of synaptic device variations to the observed degradation in accuracy by considering a comparison test case: ideal synapses with proposed neurons. That accuracy degradation amounted to 0.1\% after 35 time-steps. This implies 0.42\% degradation due to synaptic variations. 

We evaluated the energy consumption of the the basic building blocks for our system, the synaptic array and the neurons. The energy consumed by each synapse can be estimated by the transmission (or the weight) of the synaptic device. As the information being processed is based on spike events, the input can either be `1' or a `0'. Experimental demonstrations \cite{Rios_2015} have shown that readout for GST-based Si photonic devices can be achieved by pulse energies of 0.48 pJ. For our case, due to smaller GST footprints, we consider input `1' to correspond to a pulse of amplitude 0.25 mW. The power consumed by the synapse is thus given by (1-T) mW where T is the transmission of the synapse. As these read pulses will eventually write into the neurons, we choose a pulsewidth of 200 ps, which is the minimum pulsewidth required to write into the GST, as we observed previously \cite{chakraborty2018all}. Considering these metrics for the read pulses and power calculations for each synapse, we estimated the energy consumption of the entire classification operation described above. The resulting average energy consumption for first layer of the neural network in the synaptic array was calculated to be $\sim 12.5 fJ$ per synapse per time-step of evaluation. For the second layer, the energy consumption was $\sim 1.6 fJ$ per synapse per time-step. The difference is energy consumption in the two layers is due to more sparse spiking activity in the second layer. The energy consumed by each neuron was calculated in our previous work to be $5 pJ$ per time-step. The writing energies for PCM devices of similar feature sizes \cite{lee2009architecting, wong2010phase} in the electrical domain can amount upto 14-19 pJ while operating at speeds of 40-100ns. The total energy consumption for an image classification was calculated $\sim 261 nJ$ ($178 nJ$ consumed by the synaptic operations and $83 nJ$ consumed by the neurons). Although the energy consumption is comparable to CMOS technology \citep{sengupta2016probabilistic}, photonics potentially offers a faster operation at sub-ns speeds. To   note,   in   this work,   we   have   considered   a   significantly   high   read   pulse   (0.25   mW)   through   the   synapses which  is  reflected  in  the  high  energy  per  inference  operation.  The  proposed  synapses  can  be  potentially  read  with a  pulse  of  lower  amplitude  based  on  the  sensitivity  of  the  photodetectors  and  that  will  significantly  improve  the  energy requirements of the system. Moreover, the speed of operation in the photonic domain is significantly higher since read latencies  of  the  neuromorphic  systems  based  on  memristors  usually  occur  in  orders of ns.  These  benefits  encouraged us  to  further  explore  the  possibility  of  neuromorphic  hardware  design  based  on  this  technology.    
\section{Discussion}
\begin{figure}[t]
		\centering
		\includegraphics[width=3.3in, keepaspectratio]{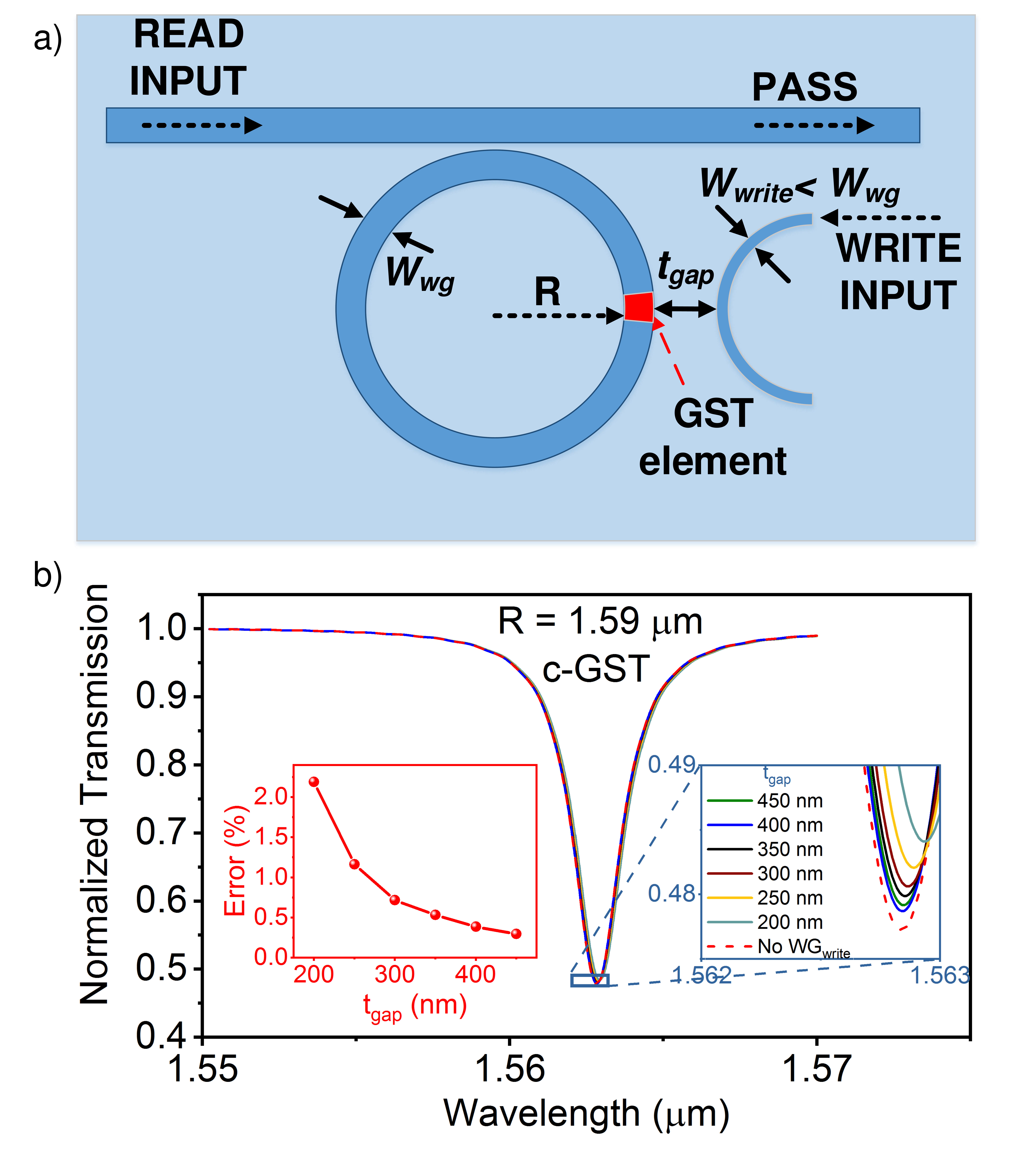}
		\caption{(a) Structure and arangement of input write waveguide at a distance $t_{gap}$ to the synaptic device. The width of the write waveguide ($W_{write}$) is smaller than that of the ring waveguide ($W_{wg}$) for asymmetric coupling. (b) Transmission characteristics of 1.59 $\mu m$ ring for different values of $t_{gap}$ compared with the case without a write waveguide. Inset 1 (Blue) shows a zoomed-in view of the transmission characteristics to show the different cases clearly. Inset 2 (Red) shows the variation of percentage error in transmission at read wavelength 1562.85 nm with $t_{gap}$.}
		\label{fig:reconf}
	\end{figure}
The proposed photonic SNN inferencing framework fills a major void of scaling from device to systems in current state-of-the-art photonic neuromorphic works based on PCMs. However, few challenges stand in the way of physical demonstration of the proposal that need to be overcome. Firstly, reconfigurability of the proposed non-volatile synaptic array is a necessity. Various reconfigurability schemes have been explored on the phase-change based photonic platforms \cite{pernice2012photonic, Zheng_2018}. We explored the possibility of adding an input bend waveguide (WG\textsubscript{write}) as a writing port for each synapse at a distance such that the inferencing framework is unaffected. The width of WG\textsubscript{write} ($W_{write}$) is intentionally considered to be much lower than the ring waveguide of the synaptic device. This is done to achieve asymmetric coupling such that during writing, the wave leaks out of WG\textsubscript{write} appropriately for efficient writing while during standard inferencing operation, the wave remains mostly confined within the ring. Fig. \ref{fig:reconf} (a) shows the structure and arrangement of WG\textsubscript{write} adjacent to the proposed synaptic device. $t_{gap}$ denotes the distance between the ring waveguide and WG\textsubscript{write}. We observe that error in transmission during normal inferencing operation due to the presence of the WG\textsubscript{write} is around 0.5 \% for $t_{gap} \sim 300 nm$. For the same distance, we calculated the transient field coupling from the WG\textsubscript{write} to the ring to be 70 \%. Thus, this writing scheme is a viable option for achieving reconfigurability in the proposed network. 

The dimensions chosen for our analysis are catered towards achieving desirable functionality for ring resonators of small radii of around $\sim 1.5 \mu m$. The main motivation behind using small ring resonators was to achieve high area density for scalability. We have explored a number of challenges arising from such small rings such as non-uniform bending and coupling losses across the range of wavelength and fabrication difficulties to achieve critical coupling. We have attempted to mitigate such challenges by appropriate design. Further, we delineated the design constraints for scaling individual synapses to a network of synapses which is necessary for large-scale neuromorphic systems. GST-based photonic platforms also experience a small resonance shift between the different programmable states of the PCM. The resonance shift between the any two states can be quantified by \cite{Stegmaier_2016}:
\begin{equation}
\frac{\Delta\lambda_m}{\lambda_{m,in}} = \frac{\Delta n_{eff,GST}}{n_{g,eff}}.\frac{L_{GST}}{2\pi R_{ring}}
\end{equation}
Here, $\lambda_{m,in}$ is the resonant wavelength in the initial state,  $\Delta n_{eff,GST}$ is the difference in effective refractive index between the states, $n_{g,eff}$ is the group index. For our case, it amounts to approximately 0.012 nm. In addition to the variations arising from device characteristics, we also explored errors arising due to interference from adjacent channels and their impact on the performance of the proposed photonic SNN. From our analysis, it can be observed that the network size, $N$ considered in our synaptic fabric is a rather conservative design. $N$ can be further increased which would result in higher errors. However, the effect of such variations have been modelled in Eqn (9) and the resulting accuracy degradation can be recovered by modifying the training algorithm as explored for memristive technologies \cite{chakraborty2018technology}. \par
The challenges of errors arising due to interference between adjacent rings essentially stems from the usage of WDM-based computation. To that effect, the limitations of array size due to WDM merits discussion. WDM, while introducing parallelism in the system, is constrained by the finesse of the rings. In this work, we have shown that we can use 16 rings in a single dot-product engine row which implies that the array can process 16 inputs in parallel. The size of the array is thus limited to $16\times N$ where N would be limited by the area and not design constraints. However, analogous computing units in the electrical domain using memristive crossbars are also limited in size due to electro-migration limits, sneak-paths and line-resistances. The photonic array on the other hand, although limited in one direction due to finesse, can be possibly extended to larger sizes in the direction of $N$. Moreover, time multiplexing is a popular practice when implementing large scale neural networks on memristive networks, as alluded to earlier. The possibility of fast writing into PCMs can potentially make these photonic arrays more suitable for temporally scalable architectures. \par
An alternative way to implement Photonic Neural Networks is through the use of inteferometers \mbox{\cite{shen2017deep}} where the weights of the network are controlled through phase-shifters. Such phase-shifters can consume significant amount of power per synapse to maintain the weight. On the other hand, non-volatile elements based on PCMs can potentially encode the weights without requiring any power to maintain their states. However, we do not use the concept of phase-shift for our design. We encode the weights in terms of levels of partial crystallization. Non-volatility is necessary for large-scale neuromorphic systems for primarily two reasons: i) it eliminates the need for phase-shifters as constant tuning is not required, and ii) it provides a platform for in-memory computing rather than storing the synaptic weights in a separate memory. In this work, the intention to use non-volatile material based memory primitive is to eliminate the need for thermal tuners. To the best of our knowledge, this is the first proposal of photonic neuromorphic platform from a scalable system point of view based on a non-volatile memory primitive. Recent proposals \mbox{\cite{shainline2018superconducting1, shainline2018superconducting2}} have looked at scalable systems to realize complex neural dynamics using for dynamic learning. However, the flux-based memory in such systems are dependent on temperature and also on the run-time of operation. Such detailed neuro-biological functionalities make them more suitable for brain-like simulations similar to NeuroGrid \mbox{\cite{benjamin2014neurogrid}} in the electrical domain. In this work, we do not incorporate complex biological dynamics of SNNs in our system and rather focus on leveraging the inherent sparsity of spike-based processing while performing image classification for energy efficiency. The primary motivation behind exploring this primitive stems from building a potentially reconfigurable neuromorphic system which performs energy-efficient inferencing. For building such neuromorphic platforms to perform spike-based processing in standard architectures, in-memory computing offers significant promise. To that effect, non-volatile memory primitives are quintessential and more suitable as they potentially eliminate the need for off-chip DRAM accesses, thus alleviating memory bottlenecks. \par

A popular way of implementing such spike-based inferencing systems is to train a network as an Artificial Neural Network (ANN) and then convert it to a Spiking Neural Network (SNN) by well explored conversion algorithms\mbox{\cite{diehl2015fast}}. This method has seen considerable success \mbox{\cite{sengupta2018going}} in image classification, far beyond the scope of spike-based training algorithms. The neurons in ANNs are usually non-linear mathematical functions, such as Rectified Linear Units (ReLU), sigmoid or tanh with ReLU being the most popularly chosen neuron functionality. During conversion, an artificial neuron with ReLU functionality can be directly converted to an IF neuron, mathematically \mbox{\cite{sengupta2018going}}. This explains why we have chosen IF neuron as the spiking neuron in our proposal. IF neurons are not associated with time-constants as it does not include leak factors and the operations are fairly simple unlike other spiking neurons. The proposal concerns with building spike-based photonic neuromorphic inferencing platform for image classification task. Note, the neuron does not bear exact resemblance to biological neuron, however, the design leverages the event-driven behavior of biological neurons. The aim of this work is to build a fast neuromorphic inferencing platform in the spiking domain to perform machine learning tasks such as image classification. Several works \mbox{\cite{benjamin2014neurogrid}} have previously explored brain-like neuron and synaptic functionalities with more significant resemblance for complex neural simulations, albeit in the electrical domain.

The major advantage of building neuromorphic systems based on Photonics rests in its speed of operation. The primary bottleneck in `write' latencies arise from the programming time of the IF neuron which can also be performed at $200 ps$. Although the current technology is power expensive during writing, the speed of writing still enables us to achieve a reasonable energy efficiency. With further optimization of switching techniques or by use of alternative PCMs with lower switching power, further energy benefits can also be aimed for to achieve comparable energy consumption to other technologies in the electrical domain. In turn, the proposed photonics computing platform eliminates various drawbacks usually faced in the electrical counterparts such as metal wire resistance, electromigration, sneak paths, etc. Despite the inherent challenges in the design and implementation, our proposed SNN framework based on GST-on-silicon photonics neuromorphic fabric enables parallelism through integration of a synaptic network with IF neurons. Such a design paves the way for scalable photonic architectures suitable for large-scale neuromorphic systems catered to perform fast computations.

\section{Conclusion}
We have proposed a photonic Spiking Neural Network computing primitive through seamless integration of non-volatile synapses and `Integrate-and-Fire' Neurons based on Phase-change materials. The microring resonator devices explored for such synapses and neurons leverage the differential optical absorption of GST for non-volatility. We use the WDM technique to scale individual synapses into a large-scale synaptic array capable of performing parallelized dot-products. Our design is based on ring resonators of radius comparable to the wavelength of operation in order to achieve high area density while maintaining performance. We explore several challenges involved in such small ring resonators and proposed certain design modifications to achieve uniform and desirable characteristics across the entire operating range of wavelength. Finally, we developed a device to system level framework to evaluate the performance of the proposed photonic in-memory computing primitive and IF neurons as an SNN inferencing engine by building behavioral models of the photonic neuromorphic fabric and achieve comparable performance to an ideal network. Neuromoprhic systems based on Integrated Photonics offer an alternative dimension to the current wave of exploring beyond von-Neumann computing frameworks and our proposed photonic SNN inferencing engine achieves a significant step towards proposing individual non-volatile devices capable of performing in-memory computing and scaling to a network of such devices to realize a truly integrated Spiking Neural Network.    

%


\section*{Acknowledgment}

The work was supported in part by, ONR-MURI program, the National Science Foundation, Intel Corporation and by the DoD Vannevar Bush Fellowship.





\bibliography{Syn}
%



%




\end{document}